\pdfoutput=1

\documentclass[11pt]{article}

\usepackage{acl}

\usepackage{times}
\usepackage{latexsym}

\usepackage[T1]{fontenc}

\usepackage[utf8]{inputenc}

\usepackage{microtype}

\title{Effective Representation to Capture Collaboration Behaviors between Explainer and User}

\author{Arjun Akula, Song-Chun Zhu \\ UCLA \\
  \texttt{aakula@ucla.edu, sczhu@stat.ucla.edu}} 

\begin{document}
\maketitle
\begin{abstract}
An explainable AI (XAI) model aims to provide transparency (in the form of justification, explanation, etc) for its predictions or actions made by it. Recently, there has been a lot of focus on building XAI models, especially to provide explanations for understanding and interpreting the predictions made by deep learning models. At UCLA, we propose a generic framework to interact with an XAI model in natural language. 
\end{abstract}

\section{Introduction}
Most work on XAI typically focuses on black-box models and generating explanations about their performance in terms of, e.g., feature visualization and attribution~\cite{sundararajan2017axiomatic,ramprasaath2016grad,zeiler2014visualizing}. However, solely generating explanations, regardless of their type (visualization or attribution) and utility, {\em is not sufficient} for increasing understandability and predictability. Current works on XAI generate explanations about their performance in terms of, e.g., feature visualization and attention maps~\cite{sundararajan2017axiomatic,ramprasaath2016grad,zeiler2014visualizing,smilkov2017smoothgrad,kim2014bayesian,zhang2018interpretable}. However, solely generating explanations, regardless of their type (visualization or attention maps) and utility, {\em is not sufficient} for increasing understandability and predictability~\cite{DBLP:journals/corr/abs-1902-10186}

In our UCLA lab, our focus is on the Explainer module. Explainer takes a natural language question from the user and identifies the intention behind it. Explainer is also responsible for controlling the dialog flow with the user. Explainable performer provides the important evidences that are necessary to answer user’s question. Atomic Performer assists Explainable performer in identifying the evidences. Explainer uses this evidence to generate most acceptable and convincing explanation. 
We control the dialog flow inside the Explainer using discourse model called Rhetorical Structure Theory (RST). In general, RST would be an efficient and simplest way to track contextual information. Since explanations are context-dependent, we believe that RST would be the right model to capture contextual information in the Explainer~\cite{DBLP:journals/corr/abs-1903-02252, akula20words, akula2019visual, akula2021crossvqa,akula2021measuring,akula2021robust}.

Given a user’s question, we first identify the dialog act of the question. We then identify the question type (contrast type) and explanation type as mentioned in the next section. Based on the explanation type, we generate the explanation and present it to the user~\cite{akula2013novel, akula2018analyzing, akula2021mind, gupta2016desire, akula2019explainable, akula2021gaining, akula2019x, akula2020words}.

Questions posed by the user to obtain explanations from an XAI model are typically contrastive in nature. For example, questions such as "Why do the model think the people are in sitting posture?", "Why do you think two persons are sitting instead of one?", need contrastive explanations. In order to generate a convincing explanation, XAI model needs to understand the implicit contrast that it presupposes~\cite{agarwal2018structured, akula2019natural, akula2015novel, palakurthi2015classification, agarwal2017automatic, dasgupta2014towards}.

Explainer’s knowledge using the question types such as NOT-X, NOT-X1-BUT-X2, NOT-X-BUT-Y. Question types such as DO-X, DO-NOT-X and DO-X-NOT-Y are used by the user as intervention techniques. We now propose the following seven types of explanation types that are motivated from an algorithmic approach rather than on theoretical grounds.
\begin{enumerate}
    \item Direct Explanation: Explaining based on detection scores
     \item Part-based Explanation: Explaining based on the evidences of detected parts for the concept asked
\item Causal Explanation, Temporal Explanation: Explaining based on the constraints from the spatiotemporal surround
\item Common-sense Explanation: Explaining based on the common-sense knowledge of the concept domain
\item Counter-factual Explanation: Explaining based on the evidences provided for the counter-factual questions asked by the Explainer
\item Discourse Entailment based Explanation: Explaining based on the discourse relations among various objects/frames in the concept/videos~\cite{akula2020cocox, r2019natural, pulijala2013web, gupta2012novel}.
\end{enumerate}

\section{Explainable AI models using Discourse relations}
Given a document (or a paragraph), discourse relation tell us how two segments (or sentences) in the document are logically connected with each other. To be more specific, discourse relations (often represented as a tree) tell us what is the function of each segment of the document, plausible reason for the presence of a segment, role of each segment w.r.t other segments, etc. In a nutshell, it provides explanation of the coherence of document/text.

In this work, we aim to propose the idea of the discourse phenomenon for videos. The intuition behind this is simple: any video can be mapped to a document (or a group of sentences). And discourse can be used to explain coherence of any document. Therefore we can use discourse to explain coherence of a video. We believe that obtaining discourse cues from videos would help us in generating better context-sensitive explanations. 

We are trying to learn the mapping from videos to its corresponding discourse representation (i.e. discourse tree) using end-to-end learning models. There are quite a few methods in the literature that try to map video/image to text. Also, there are few approaches that aim to extract hierarchical structure from the videos using unsupervised methods. However none of these approaches try to understand the discourse phenomenon in the videos.

We are currently working on two different approaches to learn the mapping from videos to its discourse representation. In the first approach, we map input videos to a sequence of vectors. We flatten the output discourse trees to a paragraph and then we use hierarchical RNN (two-levels). Basically the first level RNN captures intra-sentence dependencies and the second level RNN captures inter-sentence dependencies. In the second approach, we would like to use multi-task learning i.e. a combination of unsupervised and supervised methods. We plan on using Variational Auto-encoder as an unsupervised method and hierarchical RNN as a supervised method.

We plan on using TACoS-MultiLevel dataset: It contains 185 long videos (6 minutes on average) filmed in an indoor environment. Descriptions are manually annotated for each video. The videos are closed-domain, containing different actors, fine-grained activities, and small interacting objects in daily cooking scenarios. There are about 280 sentences in each video description on an average. Moreover, to obtain the discourse trees for the above dataset, we can use an off-the-shelf discourse tagger (like RST tagger) to map the video descriptions to discourse trees. 

We believe that an And-Or graph (AOG) representation is efficient to capture the underlying evidences used by the XAI model in making a prediction. We aim to show that the AOG representation facilitates the XAI model in generating context based explanations. We are working on building a rule-based algorithm to predict the most appropriate explanation type for a given user question. We plan to evaluate our approach on two explanation datasets: Visual Question Answering Explanation dataset (VQA-X) and Action Explanation dataset (ACT-X). We believe that the proposed question and explanation categories are sufficient enough to represent all the variations of questions and explanations in the datasets. We also plan on demonstrating that AOG representation is the key in generating most appropriate explanation for a given user question.

We plan to evaluate the effectiveness of our explanation interface and the dialogue. Extending principles identified in [Kulesza 2015], we will apply different evaluation metrics based on the following evaluation criteria~\citep{smilkov2017smoothgrad,sundararajan2017axiomatic,zeiler2014visualizing,kim2014bayesian,zhang2018interpretable}:
\begin{enumerate}
\item Correctness in recognizing the type of intents of users’ explanation requests. Different types of intention will require different types of explanations. Precision/recall/F1 measures will be applied on intention recognition. 
\item Relevancy of explanations. The provided explanation needs to satisfy the user’s intention. Accuracy will be used to measure the relevancy. We will also measure the generated explanations (e.g., language) with the ground-truth explanations using metrics such as METEOR and CIDEr [Hendricks 2016]. 
\item Fidelity to the interpretable model’s behaviors. This metric estimates how well X-pg matches STC-pg. Standard costs or similarity of matching graphs will be used for estimating fidelity. Low fidelity between the two parse graphs may be justified in some cases, e.g., when the Explainer has to summarize fragments in the STC-pg and hence avoid overwhelming the User with detailed explanations.
\end{enumerate}

\bibliography{anthology,custom}
\bibliographystyle{acl_natbib}

\end{document}